\documentclass[prc,superscriptaddress,unsortedaddress,twocolumn,showpacs,preprintnumbers,amsmath,amssymb,floatfix,dvipdfmx]{revtex4}
\usepackage[dvipdfmx]{graphicx}
\usepackage{amsmath}
\usepackage{amssymb}
\usepackage{times}
\usepackage{color}
\newcommand{\beq}{\begin{equation}}
\newcommand{\eeq}{\end{equation}}
\newcommand{\bea}{\begin{eqnarray}}
\newcommand{\eea}{\end{eqnarray}}

\newcommand{\bfi}[1]{\mbox{\boldmath $#1$}}

\newcommand{\vb}{{\bfi b}}

\newcommand{\vk}{{\bfi k}}

\newcommand{\vR}{{\bfi R}}

\def\a{\alpha}

\begin{document}

\title{
Eikonal reaction theory for two-neutron removal reactions
}

\author{K. Minomo}
\affiliation{Department of Physics, Kyushu University, Fukuoka 812-8581, Japan}

\author{T. Matsumoto}
\affiliation{Department of Physics, Kyushu University, Fukuoka 812-8581, Japan}

\author{K. Egashira}
\affiliation{Department of Physics, Kyushu University, Fukuoka 812-8581, Japan}

\author{K. Ogata}
\affiliation{Research Center for Nuclear Physics (RCNP), Osaka
University, Ibaraki 567-0047, Japan} 

\author{M. Yahiro}
\affiliation{Department of Physics, Kyushu University, Fukuoka 812-8581, Japan}

\date{\today}

\begin{abstract}
The eikonal reaction theory (ERT) proposed lately is a method of calculating 
one-neutron removal reactions at intermediate incident energies 
in which Coulomb breakup is treated accurately with 
the continuum discretized coupled-channels method. 
ERT is extended to two-neutron removal reactions. 
ERT reproduces measured one- and  two-neutron removal cross sections 
for $^6$He scattering 
on $^{12}$C and $^{208}$Pb targets at 240 MeV/nucleon and also 
on a $^{28}$Si target at 52 MeV/nucleon. 
For the heavier target in which Coulomb breakup is important, ERT 
yields much better agreement with the measured cross sections than 
the Glauber model. 
\end{abstract}


\maketitle

\textit{Introduction.}
Removal reactions are a quite useful tool for investigating 
structure of valence nucleons in weakly-bound nuclei such as 
one- and two-neutron halo nuclei. 
Spectroscopic factors and orbital angular momenta of valence nucleons 
in incident nuclei can be deduced from the removal reactions; see 
for example Ref.~\cite{Gade}.
In particular, two-neutron removal reactions are crucial for analyzing 
two-neutron correlations between valence neutrons. 
The two-neutron removal reactions were investigated for 
light targets~\cite{Ber98,ST10,ST11} with the Glauber model~\cite{Glauber}.

The Glauber model is based on the eikonal and adiabatic approximations. 
The theoretical foundation of the model is shown 
in Ref.~\cite{Yahiro-Glauber}. 
Once Coulomb breakup is taken into account 
in the Glauber model, the calculated removal cross sections 
diverge because of the adiabatic approximation. For this reason, the model 
has been applied to light targets. 
Lately a way of making Coulomb corrections to Glauber-model calculations 
was proposed~\cite{Ibrahim,Capel-08}; the divergent dipole component of 
the eikonal Coulomb phase is replaced by that estimated 
by the first-order perturbation.

The Coulomb problem is solved by the eikonal reaction theory (ERT) 
\cite{Yah11,Has11} in which Coulomb breakup is treated accurately 
with the continuum discretized coupled-channels method (CDCC)
\cite{CDCC-review1,CDCC-review2,Yah12}. 
For one-proton and -neutron removal cross sections of deuteron scattering 
at 200 MeV/nucleon, the Glauber-model results are found to be largely 
deviated from the ERT results for heavier targets \cite{Has11}. 
In Ref.~\cite{Yah11}, ERT was applied to recently measured  
one-neutron removal cross sections for $^{31}$Ne scattering from $^{12}$C and 
$^{208}$Pb targets at 230 MeV/nucleon \cite{Nak09}. 
Spectroscopic factors and asymptotic normalization coefficients for the
$^{30}$Ne~+~$n$ bound system are consistently deduced from the measured
cross sections for both the light and heavy targets. The analysis for
both light and heavy  
thus makes it possible to determine spectroscopic factors and 
asymptotic normalization coefficients of valence neutrons definitely.

Scattering of three-body projectiles such as $^{6}$He 
can be described by the four-body model composed of 
three constituents of projectile and a target. 
Four-body CDCC \cite{Matsumoto3,Matsumoto4,THO-CDCC,4body-CDCC-bin,Mat10} 
is a method of treating projectile breakup in the four-body scattering. 
Four-body CDCC was applied so far 
to elastic scattering and exclusive breakup reactions of 
$^{6}$He from $^{12}$C and $^{208}$Pb 
targets \cite{Matsumoto3,Matsumoto4,THO-CDCC,4body-CDCC-bin,Mat10}, 
$^{6}$Li elastic scattering from a $^{209}$Bi target~\cite{Watanabe:2012xs}, 
$^{11}$Li elastic scattering from a $^{209}$Bi target~\cite{Cub12}
and $^{16}$C elastic scattering from a $^{12}$C target \cite{Sasabe:2013dwa} 
with success in reproducing the experimental data, and 
importance of projectile breakup was shown for the three-body projectiles.

In this BriefReport, we extend ERT to two-neutron removal reactions. 
As a test calculation to show the validity of ERT, we analyze 
measured one- and two-neutron removal cross sections for $^6$He scattering 
on $^{12}$C and $^{208}$Pb targets at 240~MeV/nucleon 
and also on a $^{28}$Si target at 52~MeV/nucleon. 
Here $^6$He is described by the $n$~+~$n$~+~$\alpha$ 
three-body model, and $^6$He removal reactions from a target (T) is analyzed by 
four-body CDCC based on the $n$~+~$n$~+~$\alpha$~+~T model. Since 
$^6$He is well described by the three-body model~\cite{Mat10}, 
the spectroscopic factor is assumed to be 1. Hence 
the four-body CDCC calculations has no free parameter and 
$^6$He removal reaction is a good case to show the validity of ERT. 
The validity of the Glauber model is also discussed.

\textit{Formulation}. 
We start with the $n$~+~$n$~+~$\alpha$~+~T four-body system to 
analyze scattering of $^6$He on T; 
for later convenience, the two neutrons are labeled by $n_1$ and $n_2$. 
The Schr\"{o}dinger equation for the four-body system 
can be written by 
\bea
 \Big[ K_{R}+U+h-E \Big]
 \Psi&=&0 
\label{four-sch}
\eea
with 
\bea 
 U&=&U_{n_1}^{\rm (Nucl)}+U_{n_2}^{\rm (Nucl)}+
  U_{\alpha}^{\rm (Nucl)}+U_{\alpha}^{\rm (Coul)},
\label{U}
\eea
where $K_{R}$ is the kinetic energy operator with respect to 
the relative coordinate 
$\vR=(\vb,Z)$ between $^{6}$He and T, $h$ the internal Hamiltonian 
of $^{6}$He, $E$ the total energy of this system, and $\Psi$ is 
the total wave function. 
Here $U_{x}^{({\rm Nucl})}$ for $x=n_1, n_2$ and $\a$ stands for 
the nuclear part of the potential $U_{x}$ between $x$ and T, whereas 
$U_{\a}^{({\rm Coul})}$ denotes the Coulomb part of the potential $U_{\a}$ 
between $\a$ and T.

Following Ref. \cite{Yah11}, we assume the product form 
$\Psi={\hat O}\psi$ for $\Psi$. Here the operator ${\hat O}$ is defined as
\bea
\hat{O} = \frac{1}{\sqrt{\hbar \hat{v}}} e^{i\hat{K}Z} 
\eea
with the velocity operator $\hat{v}= \sqrt{2(E-h)/\mu}$ and the reduced
mass $\mu$ between $^{6}$He and T.  
Applying the eikonal approximation to Eq. \eqref{four-sch}, one 
can get the coupled-channel eikonal equation  
\bea
  i \frac{d\psi}{dZ}={\hat O^{\dagger}}U{\hat O}\psi 
\label{eikonal-eq}
\eea
for $\psi$ \cite{Yah11}. Solving Eq. \eqref{eikonal-eq} formally, one can get 
the $S$-matrix operator as \cite{Yah11}
\bea
S={\rm exp}\bigg[
-i {\cal P} \int_{-\infty}^\infty dZ \hat{O}_{}^\dagger U\hat{O}\bigg] , 
\label{S-operator}
\eea
where the \lq\lq time'' ordering operator ${\cal P}$ is 
introduced to describe multistep scattering processes accurately. 
The validity of the eikonal approximation is numerically confirmed 
for neutron and $^{4}$He elastic scattering at 50 MeV/nucleon 
by comparing full-quantum calculations with eikonal ones.

The $S$-matrix elements in the Glauber model are obtained by applying 
the adiabatic approximation to Eq. \eqref{S-operator}. 
In the approximation, $h$ is replaced by the ground-state energy $\epsilon_0$, 
and hence the operators ${\cal P}$ and ${\hat O^{\dagger}}U{\hat O}$ are 
replaced by classical numbers as ${\cal P} \to 1$ and 
${\hat O^{\dagger}}U{\hat O} \to U/(\hbar v_0)$ in Eq. \eqref{S-operator}, 
where $v_0$ is the velocity of $^{6}$He in the ground state relative to T.

At intermediate energies of our interest, the adiabatic approximation is
good for the short-range nuclear interactions $U_{x}^{\rm (Nucl)}$, but
not for the long-range Coulomb interaction $U_{\alpha}^{\rm (Coul)}$. 
In ERT, the adiabatic approximation is thus made to 
$U_{x}^{\rm (Nucl)}$ only. This leads to the following replacement:
\bea
\hat{O}_{}^\dagger U_{x}^{\rm (Nucl)} \hat{O}
\rightarrow
\frac{U_{x}^{\rm (Nucl)}}{\hbar v_0} . 
\eea
In other words, $U_{x}^{\rm (Nucl)}$ is commutable with ${\hat O}$. 
Using this property, we can separate $S$ as 
\beq
  S = S_{n_1} S_{n_2} S_\alpha
  \label{S-separation}
\eeq
with
\bea
  S_{n_1}&=&\exp\bigg[
  -\frac{i}{\hbar v_0}\int_{-\infty}^{\infty} dZ
  U_{n_1}^{\rm (Nucl)} \bigg] ,
  \label{Sn1} \\
  S_{n_2}&=&\exp\bigg[
  -\frac{i}{\hbar v_0}\int_{-\infty}^{\infty} dZ
  U_{n_2}^{\rm (Nucl)} \bigg] ,
  \label{Sn2} \\
  S_\alpha&=&\exp\bigg[
-i{\cal P}\int_{-\infty}^{\infty} dZ 
{\hat O^{\dagger}} U_\alpha {\hat O} \bigg] ,
\label{Sc}
\eea
where $U_\alpha=U_\alpha^{\rm (Nucl)}+U_\alpha^{\rm (Coul)}$. 
The operator $S_\alpha$ is the formal solution 
to the eikonal equation \eqref{eikonal-eq} with 
$U_\alpha$ instead of $U$. One can then get 
the $S$-matrix elements, 
$\langle \varphi_{0} | S_\alpha | \varphi_{0} \rangle$ and 
$\langle \varphi_{\vk} | S_\alpha | \varphi_{0} \rangle$, 
by solving Eq. \eqref{eikonal-eq} with four-body CDCC. 
The method of solving the eikonal equation \eqref{eikonal-eq} 
with CDCC was already 
formulated in Ref. \cite{Oga03} and is called eikonal-CDCC. 
The same procedure can be taken for $S_{n_1}$ and $S_{n_2}$. 
The validity of the approximation for $S_{n_1}$ and $S_{n_2}$ 
is directly confirmed by comparing four-body CDCC and
adiabatic-approximation solutions to the Schr\"{o}dinger equation
\eqref{four-sch} with no $U_{\alpha}^{\rm (Coul)}$ as shown latter.

The one- and two-neutron removal cross sections, 
$\sigma_{-1n}$ and $\sigma_{-2n}$, are described by 
\bea
\sigma_{-1n}&=&\sigma_{\rm br}^{}+\sigma_{1n~\!\rm str}~,\\
\sigma_{-2n}&=&\sigma_{\rm br}+\sigma_{1n~\!\rm str}+\sigma_{2n~\!\rm str}~ 
\label{sigma-2n}
\eea
with the elastic breakup cross section $\sigma_{br}^{}$, the one-neutron 
stripping cross section $\sigma_{1n ~\!{\rm str}}^{}$ and 
the two-neutron stripping cross section $\sigma_{2n ~\!{\rm str}}^{}$
defined by  
\bea
\sigma_{\rm br}&=&
\int d^2 \vb [
\langle \varphi_{0} | |S_\alpha S_{n_1} S_{n_2} |^2 
| \varphi_{0} \rangle 
\nonumber \\
&& ~~~~~~~~~~~~~~~~~~~~~~ - 
| \langle\varphi_{0}| S_\alpha S_{n_1} S_{n_2} 
|\varphi_{0}\rangle |^2] ,
\label{sigma-breakup}
\\
\sigma_{1n~\!\rm str}
&=&2\int d^2 \vb
\langle \varphi_{0} | |S_\alpha|^2 |S_{n_1}|^2 (1-|S_{n_2}|^2) 
| \varphi_{0} \rangle
\nonumber \\
&=& 2[\sigma_{\rm R}^{}-\sigma_{\rm br}^{}]
-2[\sigma_{\rm R}^{}(-1n)-\sigma_{\rm br}^{}(-1n)],
\\[5pt]
\sigma_{2n~\!\rm str}
&=&\int d^2 \vb
\langle \varphi_{0} | |S_\alpha|^2 (1-|S_{n_1}|^2) (1-|S_{n_2}|^2) 
| \varphi_{0} \rangle
\nonumber \\
&=& 2[\sigma_{\rm R}(-1n)-\sigma_{\rm br}(-1n)]
\nonumber \\[5pt]
&&
-[\sigma_{\rm R}-\sigma_{\rm br}]
-[\sigma_{\rm R}(-2n)-\sigma_{\rm br}(-2n)]. 
\label{two-removal-Xsec}
\eea
When $U_{\alpha}^{\rm (Coul)}=0$, these cross sections agree with those 
in the Glauber model~\cite{Ber98}; when both 
the eikonal and adiabatic approximations are taken in model calculations, 
we call the model the Glauber model for simplicity in this paper, 
even if the phenomenological optical potentials are used 
as $U_{x}^{({\rm Nucl})}$. 
Here the total reaction cross section $\sigma_{\rm R}^{}$ is defined by 
\bea
\sigma_{\rm R}^{}=
\int d^2 \vb [1-
| \langle \varphi_{0} | S_\alpha S_{n_1} S_{n_2} 
| \varphi_{0} \rangle|^2] 
\label{sigma_R}
\eea
and, $\sigma_{\rm R}$ and $\sigma_{\rm br}$ are obtained 
by solving the eikonal equation \eqref{eikonal-eq} with four-body CDCC. 
The elastic breakup and total reaction cross sections, 
$\sigma_{\rm br}^{}(-1n)$ and $\sigma_{\rm R}^{}(-1n)$,  are defined 
by Eqs. \eqref{sigma-breakup} and \eqref{sigma_R} in which 
$S_\alpha S_{n_1}S_{n_2}$ is replaced by $S_\alpha S_{n_1}$. 
Hence $\sigma_{\rm br}^{}(-1n)$ and $\sigma_{\rm R}^{}(-1n)$ are obtained by 
solving the eikonal equation \eqref{eikonal-eq} with $U_\a+U_{n_1}$ 
instead of $U$ by using four-body CDCC. 
Similarly, the elastic breakup and total reaction cross sections,
$\sigma_{\rm R}^{}(-2n)$ and $\sigma_{\rm br}^{}(-2n)$, are obtained by 
solving the eikonal equation \eqref{eikonal-eq} with $U_\a$ 
instead of $U$ by using four-body CDCC. 
All of $\sigma_{\rm br}$, $\sigma_{1n~\!\rm str}$, $\sigma_{2n~\!\rm str}$ 
and $\sigma_{-2n}$ are thus obtainable with four-body CDCC.

In actual four-body CDCC calculations, we take the same modelspace and 
internal Hamiltonian for $^6$He as 
in Ref.~\cite{Mat10}. The calculated $S$-matrix elements are 
well converged with respect to increasing the modelspace. 
Since the experimental data for high-energy $^4$He and neutron scattering are not available,
one cannot construct any phenomenological optical potentials.
In this work, 
the optical potentials $U_x^{\rm (Nucl)}$ for the 
$x$-$A$ subsystems are obtained by folding 
the Melbourne nucleon-nucleon $g$-matrix 
interaction~\cite{Amo00} with target densities
in which the proton density is determined from the electron scattering and 
the neutron distribution is assumed to have the same geometry as 
the proton one.

\begin{figure}[htbp]
\begin{center}
 \includegraphics[width=0.45\textwidth,clip]{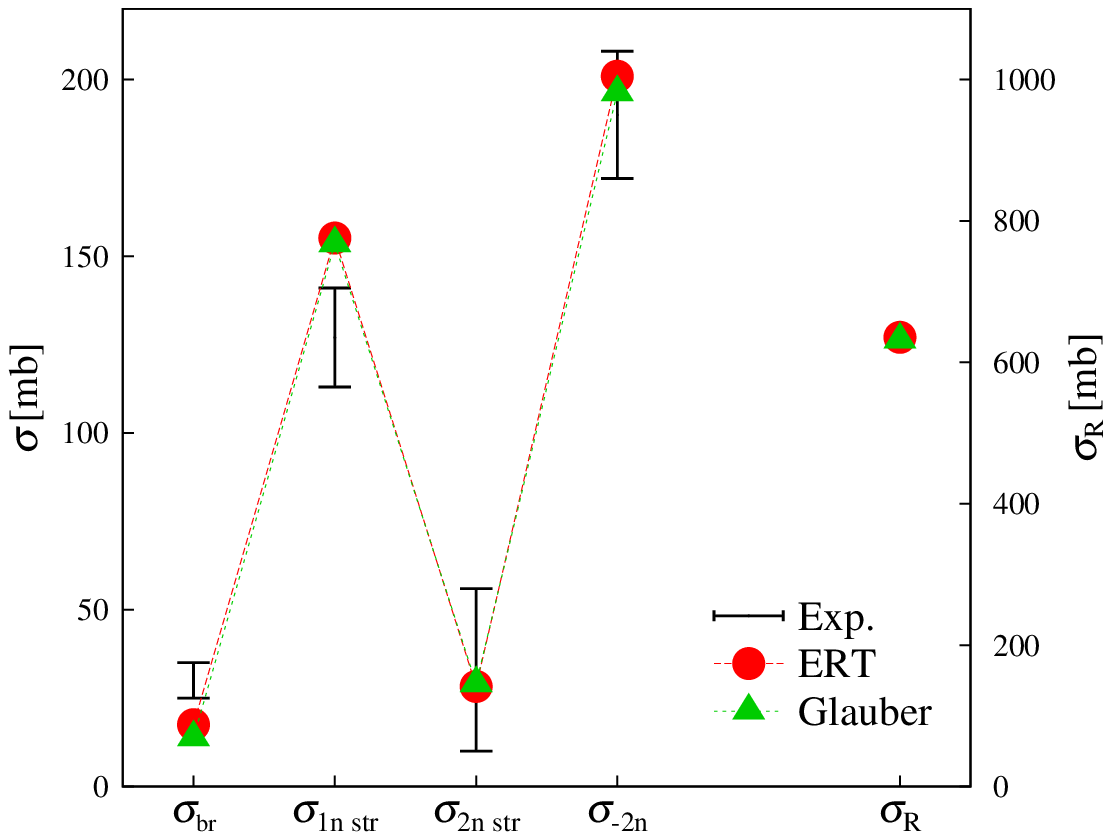}
 \includegraphics[width=0.45\textwidth,clip]{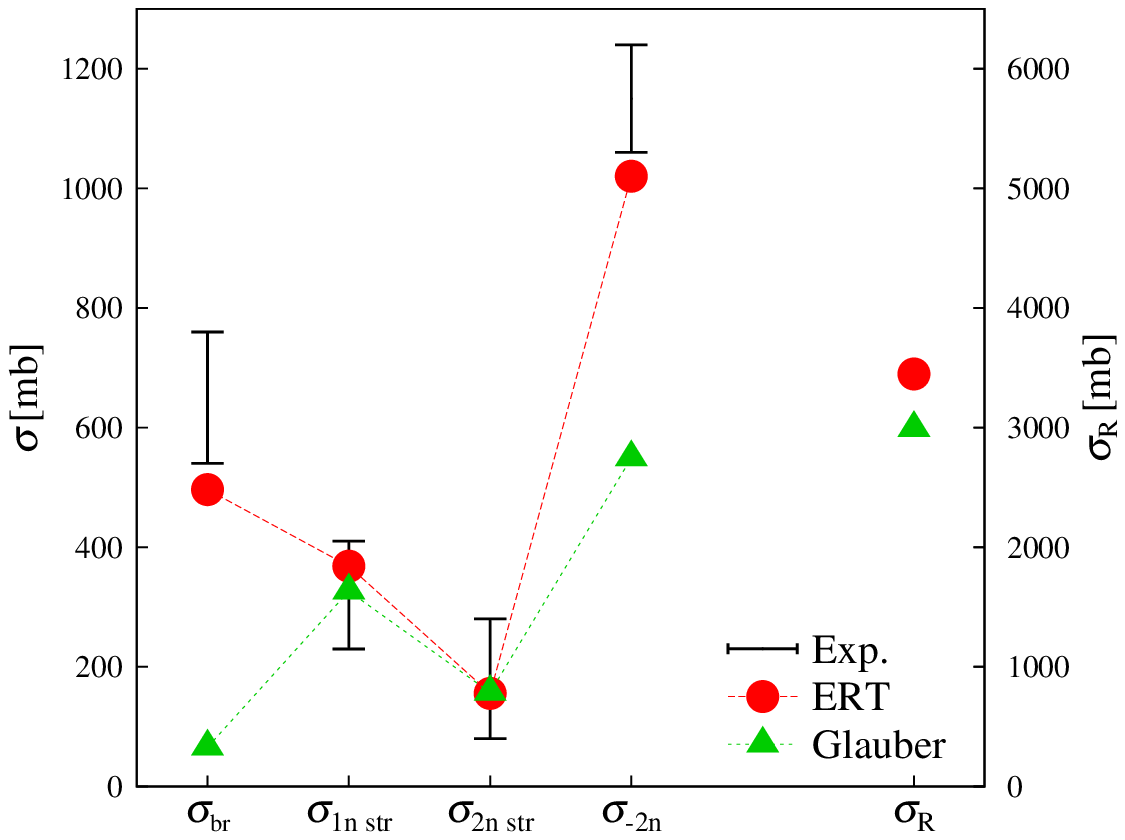}
 \caption{(Color online) 
 Elastic breakup ($\sigma_{\rm br}$), 
one-neutron stripping ($\sigma_{1n~\!\rm str}$), 
two-neutron stripping ($\sigma_{2n~\!\rm str}$), 
two-neutron removal ($\sigma_{-2n}$) and total reaction cross sections 
($\sigma_{\rm R}$) for $^6$He scattering 
from $^{12}$C (lower panel) and $^{208}$Pb (upper panel) at 240~MeV/nucleon. 
The right vertical axis stands for $\sigma_{\rm R}$, whereas 
the left one does for $\sigma_{\rm br}$, $\sigma_{1n~{\rm str}}$, 
$\sigma_{2n~{\rm str}}$, and $\sigma_{-2n}$.
The ERT and Glauber-model results are shown by 
the circles and triangles, respectively. 
Experimental data are taken from Ref.~\cite{Aum99}. 
}
 \label{fig-high-E}
\end{center}
\end{figure}

\textit{Results}. 
Figure \ref{fig-high-E} shows $\sigma_{\rm br}$, $\sigma_{1n~\!\rm str}$ 
and $\sigma_{2n~\!\rm str}$, $\sigma_{-2n}$, $\sigma_{\rm R}$ 
for $^6$He scattering from 
$^{12}$C and $^{208}$Pb targets at 240 MeV/nucleon. 
The Glauber model calculations are done by switching on the adiabatic 
approximation and off the Coulomb interaction $U_{\alpha}^{\rm (Coul)}$ 
in the ERT calculations. 
For the light target, the ERT results (solid circles) 
reproduce the experimental data~\cite{Aum99} with no free parameter, 
and the Glauber model results (solid triangles) 
are close to the ERT results. 
For the heavy target in which Coulomb breakup is important, 
the ERT results yield much better agreement with 
the experimental data than the Glauber-model results particularly 
for $\sigma_{\rm br}$ and $\sigma_{-2n}$. 
As for $\sigma_{1n~\!\rm str}$ and $\sigma_{2n~\!\rm str}$, 
the Glauber model results are close to the ERT results even 
for the heavy target, since the cross sections are determined by the absolute 
values of $S_\alpha$, $S_{n_1}$ and $S_{n_2}$ and hence mainly 
by the imaginary part of $U$. 
For the elastic breakup and two-neutron removal cross sections, meanwhile,  
the Glauber-model results underestimate the ERT ones, 
because Coulomb breakup is not included in the Glauber model. 
As a reasonable approximation, we can therefore propose the hybrid calculation 
in which $\sigma_{1n~\!\rm str}$ and $\sigma_{2n~\!\rm str}$ are calculated 
with the Glauber model and  $\sigma_{\rm br}$ with CDCC. 
The present results for $^{12}$C-target are consistent with the previous 
Glauber-model results in Ref.~\cite{Ber98}.

\begin{figure}[htbp]
\begin{center}
 \includegraphics[width=0.45\textwidth,clip]{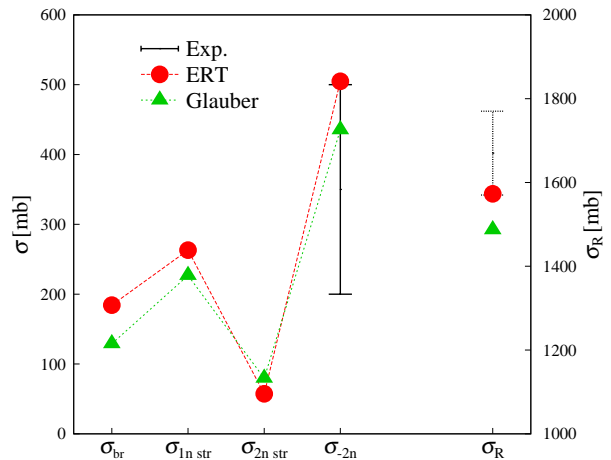}
 \caption{(Color online) 
The same as Fig.~\ref{fig-high-E} but for $^6$He scattering from 
a $^{28}$Si target at 52~MeV/nucleon. 
Experimental data are taken from Ref.~\cite{War96}. 
}
 \label{fig-low-E}
\end{center}
\end{figure}

Similar analyses are made in Fig.~\ref{fig-low-E} 
for $^6$He scattering from $^{28}$Si at 52~MeV/nucleon. 
In the analyses, the optical potential $U_\alpha^{({\rm Nucl})}$ is determined 
so as to reproduce the measured differential elastic cross section for 
$^4$He~+~$^{28}$Si scattering at 60~MeV/nucleon~\cite{You98} and the measured 
total reaction cross section at 48.1 MeV/nucl~\cite{Ing00}
by multiplying the real and imaginary parts of folding potential
by 0.91 and 1.39, respectively.
The ERT results are consistent with the experimental data for
both $\sigma_{-2n}$ and $\sigma_{\rm R}$, whereas the Glauber model
slightly underestimates the experimental data for $\sigma_{\rm R}$. 
For this incident energy, 
the deviation of the Glauber-model results from the ERT ones for 
$\sigma_{-2n}$ and $\sigma_{\rm R}$ are about 10\%, whereas 
the error of the adiabatic approximation itself is 3\% 
for $\sigma_{-2n}$ and $\sigma_{\rm R}$. The 10\% deviation is due to 
Coulomb breakup and its interference with nuclear breakup. 
The Coulomb breakup effects are more important for 
$\sigma_{\rm br}$, as expected.

\textit{Summary}. 
In this BriefReport, we extended ERT to two-neutron removal reactions. 
The method was successful in reproducing measured 
one- and two-neutron removal cross sections 
for $^6$He scattering on $^{12}$C and $^{208}$Pb targets at 240 MeV/nucleon 
and also on a $^{28}$Si target at 52 MeV/nucleon, with no free parameter. 
Particularly for the heavier target, ERT yields much better agreement
with the measured cross sections than the Glauber model. 
As a reasonable approximation, we propose the hybrid calculation 
in which $\sigma_{1n~\!\rm str}$ and $\sigma_{2n~\!\rm str}$ are calculated 
with the Glauber model and $\sigma_{\rm br}$ with CDCC.

\section*{Acknowledgements}
One of the authors (K. M.) is supported
by Grant-in-Aid for Scientific Research (No. 244137) 
from Japan Society for the Promotion of Science (JSPS).


\end{document}